%Paper: hep-ph/9205234
%From: IBANEZ%CERNVM.BITNET@pucc.princeton.edu
%Date: Tue, 26 May 92 09:29:53 SET

\input phyzzx
\input tables

\def\CLMR{\rrr\CLMR{J.A. Casas, Z. Lalak, C. Mu\~noz and
G.G. Ross, \nup347 (1990) 243;\nextline
L. Dixon, SLAC preprint 5229 (1990).}}

\def\OW{\rrr\OW{B. Ovrut and J. Wess, \plb119 (1982) 105.}}

\def\DFKZ{\rrr\DFKZ{J.P. Derendinger, S. Ferrara, C. Kounnas and
F. Zwirner,{\it ``On loop corrections to string effective field theories:
         field-dependent gauge couplings and sigma-model anomalies'',}
        preprint CERN-TH.6004/91, LPTENS 91-4                   (1991);
\nextline
G. Lopez Cardoso and B. Ovrut, \nup369 (1992) 351;\nextline
J. Louis, preprint SLAC-PUB-5527 (1991); V. Kaplunovsky and J. Louis,
as quoted in J. Louis.}}

\def\AELN{\rrr\AELN{I. Antoniadis, J. Ellis, R. Lacaze and D.V.
Nanopoulos, \plb268 (1991) 188;     S. Kalara, J.L. L\'opez and
D.V. Nanopoulos, \plb269 (1991) 84;      J. Ellis, S.Kelley and
D.V. Nanopoulos, CERN-TH.6140/91 (1991).}}

\def\FAY{\rrr\FAY{P. Fayet, \plb69 (1977) 489;\plb84 (1979) 416;
\plb78 (1978)  417;\nextline
P. Fayet and G. Farrar, \plb79 (1978) 442;\plb89 (1980) 191.}}

\def\DFS{\rrr\DFS{M. Dine, W. Fischler and M. Srednicki,
\nup189 (1981) 575;\nup202 (1982) 238;\nextline
S. Dimopoulos and S. Raby, \nup192 (1981) 353.}}

\def\MARTI{\rrr\MARTI{G. Martinelli, Review talk at the Symposium on
Lepton-Photon Interactions, Geneva 1991.}}

\def\ABJ{\rrr\ABJ{S. Adler, Phys.Rev. 177 (1969) 2426; \nextline
J.S. Bell and R. Jackiw, Nuovo Cimento 60A (1969) 47.}}

\def\WK{\rrr\WK{L. Krauss and F. Wilczek, Phys.Rev.Lett. 62 (1989) 1221.}}

\def\DIS{\rrr\DIS{T. Banks, \nup323 (1989) 90; \nextline
L. Krauss, Gen.Rel.Grav. 22 (1990) 50; \nextline
M. Alford, J. March-Russell and F. Wilczek, \nup337 (1990) 695;\nextline
J. Preskill and L. Krauss, \nup341 (1990) 50;\nextline
M. Alford, S. Coleman and J. March-Russell, preprint HUTP-90/A040 (1990).
}}

\def\WH{\rrr\WH{For a review and references see: T. Banks, Santa Cruz preprint
SCIPP 89/17 (1989).}}

\def\PR{\rrr\PR{J. Preskill, preprint CALT-68-1493 (1990).}}

\def\GRA{\rrr\GRA{R. Delbourgo and A. Salam, \plb40 (1972) 381;\nextline
T. Eguchi and P. Freund, Phys.Rev.Lett. 37 (1976) 1251; \nextline
L. Alvarez-Gaum\'e and E. Witten, \nup234 (1983) 269. }}

\def\IR{\rrr\IR{L.E. Ib\'a\~nez and G.G. Ross, \plb260 (1991) 291.}}

\def\BHR{\rrr\BHR{L. Hall and M. Suzuki, \nup231 (1984) 419;\nextline
M. Bento, L. Hall and G.G. Ross, \nup292 (1987) 400.}}

\def\ZDOS{\rrr\ZDOS{A. Font, L.E. Ib\'a\~nez and F. Quevedo, \plb2288 (1989) 79
; \nextline
A. Font, L.E. Ib\'a\~nez and  F. Quevedo  \nup345 (1990) 389.}}

\def\BRAN{\rrr\BRAN{ L. Perivolaropoulos, A. Matheson, A.C. Davis and
R. Brandenberger, preprint BROWN-HET-739 (1990).}}

\def\VIRGIN{\rrr\VIRGIN{ L.E. Ib\'a\~nez, Proceedings of the 5-th ASI on
Techniques and Concepts of High Energy Physics, St. Croix (Virgin Islands),
July 14-25, 1988. Edited by T. Ferbel, Plenum Press (1989); \nextline
A. Font, L.E. Ib\'a\~nez and F. Quevedo, \plb228 (1989) 79;\nextline
C. Geng and R. Marshak, Phys.Rev. D39 (1989) 693; \nextline
J. Minahan, P. Ramond and R. Warner, Phys.Rev. D41 (1990) 715;\nextline
R. Foot, G. Joshi, H. Lew, R. Volkas, Mod.Phys.Lett. A5 (1990) 2721.}}

\def\SS{\rrr\SS{H.P. Nilles, Phys.Rep. C110 (1984) 1;\nextline
G.G. Ross, ``Grand Unified Theories'', Benjamin Inc., (1984);\nextline
L.E. Ib\`a\~nez, ``Beyond the Standard Model (yet again)'',
CERN-TH.5982/91, to appear in the proceedings of the 1990 CERN
summer school (Mallorca).}}

\def\RPB{\rrr\RPB{L. Hall and M. Suzuki, \nup231        (1984) 419;
\nextline  F. Zwirner, \plb132 (1983) 103;\nextline
R. Mohapatra, Phys.Rev.D34 (1986) 3457;\nextline
R. Barbieri and A. Masiero, \nup267 (1986) 679;\nextline
S. Dimopoulos and L. Hall, \plb196 (1987) 135\nextline
V. Barger, G.F. Giudice and T. Han, Phys.Rev.D40  (1989) 2987.}}

\def\RP{\rrr\RP{S. Dimopoulos, R. Esmaizadeh, L. Hall and G. Starkman,
Phys.Rev.D41 (1990) 2099;\nextline
S. Dawson, \nup261 (1985) 297. }}

\def\DR{\rrr\DR{
H. Dreiner and G.G. Ross, Oxford preprint OUTP-91-15P\nextline
P. Binetruy et al., Proceedings of the ECFA Large Hadron Collider (LHC)
Workshop, Aachen, 1990. Vol.I, CERN report CERN 90-10.}}

\def\DRB{\rrr\DRB{H. Dreiner and G.G. Ross, Oxford preprint, in
preparation.}}

\def\WHH{\rrr\WHH{G. Gilbert, \nup328 (1989) 159  and references therein.
}}

\def\EEE{\rrr\EEE{J. Ellis, J. Hagelin, D. Nanopoulos and K. Tamvakis,
\plb124 (1983) 484.}}

\def\KAP{\rrr\KAP{V. Kaplunovsky, Phys.Rev.Lett. 55 (1985) 1036;
\nextline M. Dine and N. Seiberg, \plb162 (1985) 299.}}

\def\CAMP{\rrr\CAMP{B. Campbell, S. Davidson, J. Ellis and K. Olive,
\plb256 (1991) 457 ;\nextline
W. Fischler, G. Giudice, R. Leigh and S. Paban,
\plb258 (1991) 45;\nextline
H. Dreiner and G.G. Ross, in preparation. }}

\def\YAN{\rrr\YAN{M. Fukugita and T. Yanagida, \plb174 (1986) 45;
Phys.Rev.D42 (1990) 1285;
\nextline A. Bouquet and P. Salati, \nup284 (1987) 557;
\nextline  J. Harvey and M. Turner, Phys.Rev.D42 (1990) 3344;\nextline
A. Nelson and S. Barr, \plb246 (1990) 141. }}

\def\DINE{\rrr\DINE{I. Affleck and M. Dine, \nup249 (1985) 361;\nextline
D. Morgan, Texas preprint UTTG-04-91 (1991).}}

\def\HALLL{\rrr\HALLL{S. Dimopoulos and L. Hall, \plb196 (1987) 135;
\nextline J. Cline and S. Raby, Ohio preprint DOE/ER/01545-444 (1990).}}

\def\BGH{\rrr\BGH{R. Barbieri, M. Guzzo, A. Masiero and D. Tommasini,
\plb252 (1990) 251.}}

\def\PET{\rrr\PET{M. Guzzo, A. Masiero and S. Petcov,
SISSA preprint 16/91 EP (1991);   \nextline
E. Roulet, Fermilab preprint 91/18-A (1991) .}}

\def\FIQQ{\rrr\FIQQ{A. Font, L.E. Ib\`a\~nez and F. Quevedo,
\plb228 (1989) 79.}}

\def\UGO{\rrr\UGO{
J. Ellis, S. Kelley and D.V. Nanopoulos, \plb249 (1990) 441;
\plb260 (1991) 131;\nextline
P. Langacker           , "Precision tests of the standard model",
Pennsylvania preprint UPR-0435T, (1990);\nextline
U. Amaldi, W. de Boer and H. F\"urstenau, CERN-PPE/91-44 (1991);\nextline
P. Langacker and M. Luo, Pennsylvania preprint UPR-0466T, (1991).}}

\def\KAC{\rrr\KAC{ A. Font, L.E. Ib\`a\~nez and F. Quevedo,
\nup345 (1990) 389;\nextline
J. Ellis, J. L\'opez and D.V. Nanopoulos,  \plb245 (1990) 375.}}

\def\DG{\rrr\DG{S. Dimopoulos and H. Georgi, \nup193 (1981) 150.}}

\def\CHSW{\rrr\CHSW{P. Candelas, G. Horowitz,
A. Strominger and E. Witten, \nup 258 (1985) 46.}}
\def\DUAL{\rrr\DUAL{
K. Kikkawa and M. Yamasaki, \plb149 (1984) 357;
B. Sathiapalan, \prl58 (1987) 1597;
V.P. Nair, A. Shapere, A. Strominger and F. Wilczek,
      \nup287 (1987) 402;
R.~Dijkgraaf, E.~Verlinde and H.~Verlinde, \cmp115 (1988) 649;
  preprint THU-87/30;
R. Brandenberger and C. Vafa, \nup316 (1989) 391;
A. Giveon, E. Rabinovici and G. Veneziano,
    \nup322 (1989) 167;
A. Shapere and F. Wilzcek, \nup320 (1989) 669.}}

\def\FLST{\rrr\FLST{S. Ferrara,
 D. L\"ust, A. Shapere and S. Theisen,
   \plt225 (1989) 363.}}

\def\WIT{\rrr\WIT{E. Witten, \plb155 (1985) 151. }}

\def\CFGVP{\rrr\CFGVP{E. Cremmer, S. Ferrara, L. Girardello and
A. Van Proeyen, \nup212 (1983) 413. }}
\def\SH{\rrr\SH{H. Georgi, \nup156 (1979) 126. }}
\def\ADBEL{\rrr\ADBEL{S. Adler, Phys.Rev. 177 (1969) 2426. \nextline
J.S. Bell and R. Jackiw, Nuovo Cimento 60A (1969) 47.}}
\def\SAL{\rrr\SAL{R. Delbourgo and A. Salam, \plb40 (1972) 381;
\nextline T. Eguchi and P. Freund, Phys.Rev.Lett. 37 (1976) 1251.}}
\def\ALWI{\rrr\ALWI{L. Alvarez-Gaum\'e and E. Witten, \nup234 (1983) 269.
}}
\def\ASI{\rrr\ASI{L. E. Ib\' a\~ nez, Proceedings of the 5-th ASI
on Techniques and Concepts of High Energy Physics, St. Croix (Virgin
Islands), July 14-25, 1988. Edited by T. Ferbel, Plenum Press (1989).
}}
\def\FIQUNO{\rrr\FIQUNO{A. Font, L.E. Ib\' a\~ nez and F. Quevedo,
Phys.Lett. B228 (1989) 79. }}
\def\MARS{\rrr\MARS{C. Geng and R. Marshak, Phys.Rev. D39 (1989) 693;
\nextline
J. Minahan, P. Ramond and R. Warner, Phys.Rev. D41 (1990) 715;\nextline
K. Babu and R. Mohapatra, Phys.Rev. D41 (1990) 271. }}
\def\PDG{\rrr\PDG{Particle Data Group, Phys.Lett. B204 (1988) 1. }}
\def\TOOFT{\rrr\TOOFT{G. 't Hooft, in "Recent Developments in Gauge
Theories", ed. by G. 't Hooft et al., Plenum Press, New York (1981).}}
\def\CP{\rrr\CP{ For reviews of the strong-CP problem see: \nextline
J. E. Kim, Phys.Rep. 150 (1987) 1; \nextline
H. Y. Cheng, Phys.Rep. 158 (1988) 1.}}
\def\ALT{\rrr\ALT{I. S. Altarev et al., JETP Lett. 44 (1986) 461.}}
\def\SMIT{\rrr\SMIT{K. M. Smith et al., Phys. Lett. B234 (1990) 191. }}
\def\PQ{\rrr\PQ{R. Peccei and H. Quinn, Phys.RevLett. 38 (1977) 1440;
Phys.Rev. D16 (1977) 1791. }}
\def\WEWI{\rrr\WEWI{S. Weinberg, Phys.Rev.Lett. 40 (1978) 223;\nextline
F. Wilczek, Phys.Rev.Lett. 40 (1978) 279.}}
\def\INVIS{\rrr\INVIS{J. E. Kim, Phys.Rev.Lett. 43 (1979) 103;\nextline
M. Shifman, A. Vainstein and V. Zakharov, \nup116 (1980) 493;\nextline
M. Dine, W. Fischler and M. Srednicki, \plb104 (1981) 99. }}
\def\DICUS{\rrr\DICUS{D. Dicus et al., \prv22 (1980) 839;\nextline
M. Fukugita, S. Watamura and M. Yoshimura, \prv26 (1982) 1840;\nextline
A. Pantziris and K. Kang, \prv33 (1986) 3509.}}
\def\SN{\rrr\SN{R. Mayle, J. Wilson, J. Ellis and K. Olive,
\plb219 (1989) 515.}}
\def\PRESK{\rrr\PRESK{J. Preskill, M. Wise and F. Wilczek, \plb120 (1983)
127;\nextline
L. Abbott and P. Sikivie, \plb120 (1983) 133;\nextline
M. Dine and W. Fischler, \plb120 (1983) 137.}}
\def\NELS{\rrr\NELS{A. Nelson, \plb136 (1984) 387; \plb143 (1984)
165;\nextline
S. Barr, Phys.Rev.Lett. 53 (1984) 329;\prv30 (1984) 1805;\nextline
S. Barr and A. Masiero, \prv38 (1988) 366.}}
\def\GILD{\rrr\GILD{E. Gildener, \prv14 (1976) 1667.}}
\def\COSM{\rrr\COSM{M. Veltman, Phys.Rev.Lett. 34 (1975) 77 ;\nextline
A. Linde, JETP.Lett. 19 (1974) 183. }}
\def\TECH{\rrr\TECH{S. Weinberg, \prv13 (1976) 974;\prv19 (1979) 1277;
\nextline
L. Susskind, \prv20 (1979) 2619;\nextline
E. Farhi and L. Susskind, Phys.Rep. 74C (1981) 2777.}}
\def\DIMSU{\rrr\DIMSU{S. Dimopoulos and L. Susskind, \nup155 (1979) 237;
E. Eichten and K. Lane, \plb90 (1980) 125.}}
\def\DIMEL{\rrr\DIMEL{S. Dimopoulos and J. Ellis, \nup182 (1981) 505.}}
\def\HOLDO{\rrr\HOLDO{B. Holdom, \prv24 (1981) 1441;\plb150 (1985) 301;
\nextline
K. Yamawaki, M. Bando and K. Matumoto, Phys.Rev.Lett. 56 (1986) 1335;
\nextline
T. Appelquist, D. Karabali and L. Wijewardhana, Phys.Rev.Lett. 57 (1986)
957;\nextline
T. Appelquist and L. Wijewardhana, \prv36 (1987) 568;\nextline
S. Raby and G. Giudice, Ohio preprint DOE/ER/01545-447 (1990).}}
\def\NAMB{\rrr\NAMB{Y. Nambu, in New Theories in Physics, proceedings
of the XIth Kazimierz Symposium, (1988). World Scientific (1989).}}
\def\LINDN{\rrr\LINDN{W. Bardeen, C. Hill and M. Lindner,
\prv41 (1990) 1647;\nextline
V. Miransky, M. Tanabashi and K. Yamawaki, Mod.Phys.Lett. A4 (1989) 1043;
\plb221 (1989) 177. }}
\def\CDFMT{\rrr\CDFMT{CDF Collaboration, Fermilab preprint Conf-90/138-E
\nextline (1990).}}
\def\FERN{\rrr\FERN{E. Fernandez, CERN-PPE/90-151, talk given at the
Neutrino-90 Conference, CERN, June 1990.}}
\def\NAJL{\rrr\NAJL{Y. Nambu and G. Jona-Lasinio, Phys. Rev. 122 (1961)
345.}}
\def\CDFLC{\rrr\CDFLC{CDF Collaboration (presented by S. Bertolucci),
Fermilab-Conf-90/45-T (1990). Proceedings of the 8th Topical
Workshop on $p-{\bar p}$ Collider Physics, Castiglione (1989).}}
\def\CHAGA{\rrr\CHAGA{M. Chanowitz and M. K. Gaillard, \nup261 (1985)
379;\nextline
M. Chanowitz, M. Golden and H. Georgi, \prv36 (1987) 149; Phys.Rev.Lett.
57 (1986) 2344.}}
\def\DOHE{\rrr\DOHE{A. Dobado and M. Herrero, \plb228 (1989) 495;
\plb233 (1989) 505;\nextline
J. F. Donaghue and C. Ramirez, \plb234 (1990) 361.}}
\def\ESPR{\rrr\ESPR{A. Dobado, D. Espriu and M. Herrero,
CERN-TH.5785/90 (1990).}}
\def\PT{\rrr\PT{M. Peskin and T. Takeuchi, SLAC-PUB-5272 (1990).}}
\def\RAND{\rrr\RAND{M. Golden and L. Randall, Fermilab preprint
FERMILAB-PUB-90/83-T (1990).}}
\def\MARCI{\rrr\MARCI{W. Marciano and J. Rosner, BNL-44997 (1990).}}
\def\RING{\rrr\RING{A. Ringwald, \nup330 (1990) 1.}}
\def\ESPI{\rrr\ESPI{O. Espinosa, Caltech preprint CALT-68-1586 (1989).}}
\def\TOOFTB{\rrr\TOOFTB{G. 't Hooft, Phys.Rev.Lett. 37 (1976) 8;
\prv14 (1976) 3422.}}
\def\RUBA{\rrr\RUBA{V. Kuzmin, V. Rubakov and M. Shaposnikov,
\plb155 (1985) 36.}}
\def\MANTO{\rrr\MANTO{F. Klinkhamer and N. Manton, \prv30 (1984) 2212.}}
\def\SUCIN{\rrr\SUCIN{H. Georgi and S. L. Glashow, Phys.Rev.Lett. 32
(1974) 32.}}
\def\PASA{\rrr\PASA{J. Pati and A. Salam, \prv10 (1974) 275.}}
\def\GQW{\rrr\GQW{H. Georgi, H. Quinn and S. Weinberg, Phys.Rev.Lett.
33 (1974) 451.}}
\def\GUTS{\rrr\GUTS{For a review on Grund Unified Theories see:
\nextline
G.G. Ross, "Grand Unified Theories", Benjamin Inc. (1984);
\nextline
P. Langacker, Phys.Rep. 72C (1981) 185.}}
\def\PDEC{\rrr\PDEC{IMB Collaboration, S. Seidel et al. Phys.Rev.Lett.
61 (1988) 2522;\nextline
Kamiokande-II Collaboration, K. Hirata et al. \plb220 (1989) 308;
\nextline
Ch. Berger et al., \nup313 (1989) 509;\nextline
T. Philips et al., \plb224 (1989) 348.}}
\def\MARSI{\rrr\MARSI{W. Marciano and A. Sirlin, \prv22 (1980) 2095;
\nup189 (1981) 442;\nextline
C. H. Llewellyn-Smith, G. G. Ross and J. Wheater , \nup177 (1981) 263.}}
\def\BEGN{\rrr\BEGN{A. Buras, J. Ellis, M. K. Gaillard and D. Nanopoulos,
\nup195 (1978) 66.}}
\def\SODI{\rrr\SODI{H. Georgi, unpublished;\nextline
H. Fritzsch and P. Minkowski, Ann.Phys. 93 (1975) 193;\nextline
H. Georgi and D. Nanopoulos, \nup159 (1979) 59.}}
\def\SOAI{\rrr\SOAI{H. Georgi and D. Nanopoulos, \nup159 (1979) 16;
\nextline
F. del Aguila and L. E. Ib\' a\~ nez, \nup177 (1981) 60. }}
\def\SAKH{\rrr\SAKH{A.D. Sakharov, Pisma ZhETF, 5 (1967) 32.}}
\def\KUZ{\rrr\KUZ{V. A. Kuzmin, Pisma ZhETF, 13 (1970) 335.}}
\def\OKUN{\rrr\OKUN{L. Okun and Y. Zeldovich, Comm.Nucl.Part.Phys.
6 (1976) 69.}}
\def\IGNAT{\rrr\IGNAT{A. Y. Ignatiev, N. Krasnikov, V. Kuzmin
and A. Takhelidze, \plb76 (1978) 436;\nextline
M. Yoshimura, Phys.Rev.Lett. 41 (1978) 281;\nextline
S. Dimopoulos and L. Susskind, \prv18 (1978) 4500;\nextline
S. Weinberg, Phys.Rev.Lett. 42 (1979) 850.}}
\def\ARNO{\rrr\ARNO{P. Arnold and L. McLerran, \prv37 (1988) 1020;
\nextline
S. Khlebnikov and M. Shaposnikov, \nup308 (1988) 885;\nextline
D. Grigoriev, V. Rubakov and M. Shaposnikov, \plb216 (1989) 172.}}
\def\MAJOR{\rrr\MAJOR{G. Gelmini and M. Roncadelli, \plb99 (1981) 411;
\nextline
H. Georgi et al., \nup193 (1983) 297.}}
\def\VALL{\rrr\VALL{ For a review see e.g. J. Valle, Valencia preprint
FTUV/90-36, to appear in Progress in Particle and Nuclear Physics.}}
\def\RAMO{\rrr\RAMO{M. Gell-Mann, P. Ramond and R. Slansky, in
Supergravity, eds. P. van Nieuwenhuizen and D. Freedman
(North-Holland 1979) p. 315; \nextline
T. Yanagida, Proceedings of the Workshop on Unified Theories and the
Baryon Number of the Universe, KEK, Japan (1979).}}
\def\PONT{\rrr\PONT{B. Pontecorvo, Sov.Phys. JETP 6 (1958) 429;
ibid. 7 (1958) 172.}}
\def\OSCIL{\rrr\OSCIL{L. Moscoso, Saclay preprint DPhPE 90-14 (1990),
to appear in the Proceedings of "Neutrino 90", CERN, June 1990.}}
\def\DAVI{\rrr\DAVI{R. Davis, D. Harmer and K. Hoffman, Phys.Rev.Lett.
20 (1968) 1205;\nextline
J. Rowley, B. Cleveland and R. Davis, in "Solar Neutrinos and
Neutrino Astronomy, edited by M. L. Cherry;\nextline
K. Lande, "Neutrino 90", Proceedings of the 14th Int. Conf. on
Neutrino Physics and Astrophysics. Editor K. Winter
(North-Holland), in press.}}
\def\KAMIO{\rrr\KAMIO{K. S. Hirata et al. Phys.Rev.Lett. 63 (1989) 16;
Phys.Rev.Lett. 65 (1990) 1301; KEK preprint 90-43 (1990);\nextline
A. Suzuki, KEK preprint 9-31 (1990).}}
\def\SSM{\rrr\SSM{J. N. Bahcall and R. K. Ulrich, Rev.Mod.Phys.
60 (1988) 297.}}
\def\MSW{\rrr\MSW{S. Mikheyev and A. Smirnov, Sov.J.Nucl.Phys. 42
(1985) 913;\nextline
L. Wolfenstein, \prv17 (1978) 2369; \prv20 (1979) 2634;\nextline
For a review see:\nextline
S. Mikheyev and A. Smirnov, Sov.Phys.Usp. 30 (1987) 759.}}
\def\GALL{\rrr\GALL{V.N. Gavrin in "Neutrino 90", CERN, June 1990;
\nextline
T. Kirsten, ibid.}}
\def\CISN{\rrr\CISN{A. Cisneros, Astrophys.Space Sci. 10 (1981) 87;
\nextline
L. Okun, M. Voloshin and M. Vysotsky, Sov.J.Nucl.Phys. 91 (1986) 754;
Sov. Phys. JETP 64 (1986) 446.}}
\def\BABU{\rrr\BABU{K. Babu and V. Mathur, \plb196 (1987) 218;\nextline
M. Fukugita and T. Yanagida, Phys.Rev.Lett. (1987) 1807.}}
\def\SUSY{\rrr\SUSY{Y. Golfand and E. Likhtman, JETP Lett. 13 (1971) 323;
\nextline
D. Volkov and V. Akulov, Pis'ma Zh.ETF 16 (1972) 621;\nextline
P. Ramond, \prv3 (1971) 2415.}}
\def\WZ{\rrr\WZ{J. Wess and B. Zumino, \nup70 (1974) 139.}}
\def\NILL{\rrr\NILL{For phenomenological discussions of supersymmetry
see:\nextline
H. P. Nilles, Phys.Rep. C110 (1984) 1;\nextline
H. Haber and G. Kane, Phys.Rep. C117 (1985) 75;\nextline
G.G. Ross, `Grand Unified Theories' (Benjamin, New York, 1984);\nextline
S.Ferrara ed., `Supersymmetry' (2 Vols.), North-Holland-World
Scientific, Singapore (1987).}}
\def\SUHY{\rrr\SUHY{M. Veltman, Acta Phys.Polon. B12 (1981) 437;
\nextline
L. Maiani, Proceedings of the Summer School of Gif-Sur-Yvette
(Paris 1980).}}
\def\FAY{\rrr\FAY{P. Fayet, \plb69 (1977) 489;\plb84 (1979) 416;
\plb78 (1978) 417;\nextline
G. Farrar and P. Fayet, \plb79 (1978) 442;\plb89 (1980) 191.}}
\def\DRW{\rrr\DRW{S. Dimopoulos, S. Raby and F. Wilczek, \prv24 (1981)
1681.}}
\def\IRA{\rrr\IRA{L. E. Ib\' a\~nez and G. G. Ross, \plb105 (1982) 439;
\nextline
M. Einhorn and D. R. T. Jones, \nup196 (1982) 475.}}
\def\LANGS{\rrr\LANGS{P. Langacker, Pennsylvania preprint UPR-0435T
(1990).}}
\def\WEIN{\rrr\WEIN{S. Weinberg, \prv26 (1982) 287;\nextline
N. Sakai and T. Yanagida, \nup197 (1982) 533.}}
\def\SPD{\rrr\SPD{S. Dimopoulos, S. Raby and F. Wilczek, \plb112
(1982) 133;\nextline
J. Ellis, D. Nanopoulos and S. Rudaz, \nup202 (1982) 43.}}
\def\CFGVP{\rrr\CFGVP{E. Cremmer, S. Ferrara, L. Girardello and A.
Van Proeyen, \nup212 (1983) 413.}}
\def\IBA{\rrr\IBA{L. E. Ib\'a\~nez, \plb118 (1982) 73; \nup218
(1983) 514.}}
\def\BFS{\rrr\BFS{R. Barbieri, S. Ferrara and C. Savoy, \plb119
(1982) 343;\nextline
P. Nath, R. Arnowitt and A. Chamseddine, \plb49 (1982) 970.}}
\def\IBB{\rrr\IBB{L. E. Ib\'a\~nez, \nup218 (1983) 514.}}
\def\IBLO{\rrr\IBLO{L. E. Ib\' a\~ nez and C. L\' opez, \plb126
(1983) 54; \nup233 (1984) 511.}}
\def\AGPW{\rrr\AGPW{L. Alvarez-Gaum\' e, J. Polchinsky and M. Wise,
\nup221 (1983) 495.}}
\def\EHNT{\rrr\EHNT{J. Ellis, J. Hagelin, D. Nanopoulos and K. Tamvakis,
\plb125 (1983) 275.}}
\def\HPNI{\rrr\HPNI{H. P. Nilles, \plb115 (1982) 193.}}
\def\NSW{\rrr\NSW{H. P. Nilles, M. Srednicki and D. Wyler,
\plb120 (1983) 275.}}
\def\GIM{\rrr\GIM{J. Ellis and D. Nanopoulos, \plb110 (1982) 211;
\nextline
R. Barbieri and R. Gatto, \plb110 (1982) 211;\nextline
T. Inami and C. S. Lim, \nup207 (1982) 593.}}
\def\IRB{\rrr\IRB{L. E. Ib\' a\~ nez and G. G. Ross, \plb110 (1982)
215.}}
\def\SNEU{\rrr\SNEU{L. E. Ib\'a\~nez, \plb137 (1984) 160;\nextline
J. Hagelin, G. Kane and S. Raby, \nup241 (1984) 638.}}
\def\INOU{\rrr\INOU{K. Inoue et al., Prog.Theor.Phys. 67 (1982) 1859.}}
\def\INOUB{\rrr\INOUB{K. Inoue et al., Prog.Theor.Phys. 68 (1982) 927.}}
\def\ILM{\rrr\ILM{L. E. Ib\'a\~nez, C. L\'opez and C. Mu\~ noz,
\nup256 (1985) 218.}}
\def\JRR{\rrr\JRR{S. Jones and G. G. Ross, \plb155 (1984) 69;\nextline
C. Kounnas, A. Lahanas, D. Nanopoulos and M. Quiros,
\nup236 (1984) 438;\nextline
A. Bouquet, J. Kaplan and C. Savoy, \nup262 (1985) 299.}}
\def\FLSH{\rrr\FLSH{R. Flores and M. Sher, Ann.Phys. 148 (1983) 95;
\nextline
H. P. Nilles and M. Nussbaumer, \plb145 (1984) 73;\nextline
P. Majumdar and P. Roy, \prv30 (1984) 2432.}}
\def\HAHE{\rrr\HAHE{H. E. Haber and R. Hempfling, Phys.Rev.Lett. 66
(1991) 1815       ;\nextline
J. Ellis, G. Ridolfi and F. Zwirner, \plb257 (1991) 83; \plb262 (1991)
477;\nextline
Y. Okada, M. Yamaguchi and T. Yanagida, Prog.Theor.Phys. Lett. 85 (1991)
1;\plb262 (1991) 54;\nextline
R. Barbieri and M. Frigeni, \plb258 (1991) 395;\nextline
R. Barbieri, F. Caravaglios and M. Frigeni, \plb258 (1991) 167;\nextline
J.R. Espinosa and M. Quiros, \plb266 (1991) 389.}}

\def\EIR{\rrr\EIR{J. Ellis and G. G. Ross, \plb117 (1982) 397;
\nextline
J. Ellis, L. E. Ib\' a\~ nez and G. G. Ross, \nup221 (1983) 445;
\nextline
A. Chamseddine, P. Nath and R. Arnowitt, \plb129 (1983) 445;
\nextline
P. Dicus, S. Nandi and X. Tata, \plb129 (1983) 451.}}
\def\CDFMT{\rrr\CDFMT{CDF Collaboration (presented by G. P. Yeh),
Fermilab-Conf-90/138-E \nextline (1990).}}
\def\CASCA{\rrr\CASCA{H. Baer, X. Tata and J. Woodside,
Phys.Rev.Lett. 63 (1989) 352;\prv41 (1990) 906;\nextline
H. Baer, D. Karatas and X. Tata, Florida State University
preprint FSU-HEP-900430.}}
\def\ALBA{\rrr\ALBA{C. Albajar, C. Fuglesang, S. Hellman, F. Pauss
and G. Polesello, Proceedings of the LHC-Workshop, Aachen (1990).
CERN 90-10 (1990).}}
\def\SFER{\rrr\SFER{L3 Collaboration, B. Adeva et al.,\plb233 (1989) 530;
\nextline
ALEPH Collaboration, D. Decamp et al., \plb236 (1990) 86;\nextline
OPAL Collaboration, M. Z. Akrawy et al., \plb240 (1990) 261;\nextline
DELPHI Collaboration, P. Abreu et al., \plb247 (1990) 157.}}
\def\DELPHI{\rrr\DELPHI{DELPHI Collaboration, P. Abreu et al.,
CERN-EP/90-80 (1990).}}
\def\NEUT{\rrr\NEUT{ALEPH Collaboration, D. Decamp et al.,
CERN-EP/90-63 (1990);\nextline
DELPHI Collaboration, P. Abreu et al., CERN-EP/90-80;\nextline
OPAL Collaboration, M. Z. Akrawy et al., CERN-PPE/90-95;\nextline
MARK-II Collaboration, S. Komamiya et al., Phys.Rev.Lett. 64 (1990)
2984.}}
\def\BGGR{\rrr\BGGR{R. Barbieri, G. Gamberini, G. Giudice and
G. Ridolfi, \plb195 (1987) 500; \nup296 (1988) 75;\nextline
A. Bartl et al., preprints HEPHY-PUB 526/89 and UWThPh-1989-38\nextline
(1989);\nextline
J. Ellis, G. Ridolfi and F. Zwirner, \plb237 (1990) 423.}}
\def\SHIG{\rrr\SHIG{ALEPH Collaboration, D. Decamp et al., \plb237
(1990) 291;\nextline
DELPHI Collaboration, P. Abreu et al., \plb245 (1990) 276;\nextline
OPAL Collaboration, M. Z. Akrawy et al., CERN-EP/90-100 (1990);
\nextline
L3 Collaboration, B. Adeva et al., CERN-PPE-L3-015 (1990);\nextline
MARK-II Collaboration, S. Komamiya et al. Phys.Rev.Lett. 64 (1990)
2881.}}
\def\HHG{\rrr\HHG{For a useful "Higgs-hunting guide" see:
S. Dawson, J. Gunion, H. Haber and G. Kane, BNL-41644 (1989)
(to appear in Phys.Rep.).}}
\def\IM{\rrr\IM{L. E. Ib\' a\~ nez and J. Mas, \nup286 (1987) 107.}}
\def\DS{\rrr\DS{J. P. Derendinger and C. Savoy, \nup237 (1984) 307.}}
\def\DREE{\rrr\DREE{M. Drees, Int.Jour.Mod.Phys.A 4 (1989) 3635;
\nextline
J. Ellis, J. Gunion, H. Haber, L. Roszkowski and F. Zwirner,
Phys.Rev. D39 (1989) 844;\nextline
P. Binetruy and C. Savoy, Saclay preprint SPhT/91-143.}}

\def\HALL{\rrr\HALL{L. Hall and M. Suzuki, \nup231 (1984) 419;
\nextline
I. Lee, \nup246 (1984)) 120.}}
\def\ZWI{\rrr\ZWI{F. Zwirner, \plb132 (1983) 103;\nextline
R. Barbieri and A. Masiero, \nup267 (1986) 679.}}
\def\DIMH{\rrr\DIMH{S. Dimopoulos and L. Hall, \plb196 (1987) 135;
\plb207 (1988) 216;\nextline
S. Dimopoulos et al. \prv41 (1990) 2099.}}
\def\HARA{\rrr\HARA{L. Hall and L. Randall, LBL-28879 (1990).}}
\def\BG{\rrr\BG{R. Barbieri and G. Giudice, \nup296 (1988) 75.}}
\def\GSW{\rrr\GSW{M. Green, J. Schwarz and E. Witten, "Superstring
Theory", Vols I and II, Cambridge University Press (1986) ;\nextline
D. Gross, in Proceedings of the 1986 ASI School (Virgin Islands),
Plenum Press (1987).}}
\def\GS{\rrr\GS{M. Green and J. Schwarz, \nup255 (1985) 93.}}
\def\HETE{\rrr\HETE{D. Gross, J. Harvey, E. Martinec and R. Rohm,
\nup256 (1985); \nup267 (1986) 75.}}
\def\ORBI{\rrr\ORBI{L. Dixon, J. Harvey, C. Vafa and E. Witten,
\nup261 (1985) 651;\nextline
L. E. Ib\' a\~nez, H. P. Nilles and F. Quevedo, \plb187 (1987) 25;
\nextline
K. Narain, M. Sarmadi and C. Vafa, \nup288 (1987) 951;\nextline
L. E. Ib\' a\~ nez, J. Mas, H. P. Nilles and F. Quevedo,
\nup301 (1988) 157.}}
\def\KLT{\rrr\KLT{H. Kawai, D. Lewellen and S. Tye, Phys.Rev.Lett.
57 (1986) 1832; \nup288 (1987 )1;\nextline
I. Antoniadis, C. Bachas and C. Kounnas, \nup289 (1987) 87.}}
\def\LLS{\rrr\LLS{W. Lerche, D. L\" ust and A. N. Schellekens,
\nup287 (1987) 477.}}
\def\GEP{\rrr\GEP{D. Gepner, \nup296 (1987) 757;\nextline
Y. Kazama and H. Suzuki, \nup321 (1989) 232.}}
\def\TWIS{\rrr\TWIS{A. Font, L. E. Ib\' a\~nez, F. Quevedo and
A. Sierra, \nup337 (1990) 119.}}
\def\GKMR{\rrr\GKMR{B. Greene, K. Kirklin, P. Miron and G. G. Ross,
\nup278 (1986) 667; \nup279 (1986) 574.}}
\def\FENO{\rrr\FENO{
A. Font, L. E. Ib\' a\~ nez, F. Quevedo and A. Sierra, \nup331 (1990)
421.}}
\def\FLIP{\rrr\FLIP{I. Antoniadis, J. Ellis, J. Hagelin and D. Nanopoulos
, \plb231 (1989) 65 and references therein.}}
\def\GINS{\rrr\GINS{P. Ginsparg, \plb197 (1987) 139.}}
\def\SHE{\rrr\SHE{A. N. Schellekens, \plb237 (1990) 363.}}
\def\HOSO{\rrr\HOSO{Y. Hosotani, \plb129 (1983) 193;\nextline
E. Witten, \nup327 (1989) 673.}}
\def\LLEW{\rrr\LLEW{D. Lewellen, \nup337 (1990) 61;\nextline
J.A. Schwarz, Phys.Rev.D42 (1990) 389.}}
\def\LEVEL{\rrr\LEVEL{ A. Font, L. E. Ib\' a\~ nez and F. Quevedo,
\nup345 (1990) 389.}}
\def\MISS{\rrr\MISS{S. Dimopoulos and F. Wilczek, Santa Barbara
preprint (1981); Proc. Erice Summer School (1981);\nextline
B. Grinstein, \nup206 (1982) 387;\nextline
A. Masiero, D. Nanopoulos, K. Tamvakis and T. Yanagida, \plb115
(1982) 380.}}
\def\BD{\rrr\BD{T. Banks and L. Dixon, \nup307 (1988) 93.}}
\def\DIN{\rrr\DIN{J. P. Derendinger, L. E. Ib\' a\~ nez and H. P. Nilles,
\plb155 (1985) 65;\nextline
M. Dine, R. Rohm, N. Seiberg and E. Witten, \plb156 (1985) 55.}}
\def\ANOM{\rrr\ANOM{M. Green and J. Schwarz, \plb149 (1984) 117.}}

\def\FDGS{\rrr\FDGS{E. Witten , \plb149 (1984);\nextline
W. Lerche, B. Nilsson, A.N. Schellekens, \nup299 (1988) 91;\nextline
M. Dine, N. Seiberg and E. Witten, \nup289 (1987) 585;\nextline
J. Atick, L. Dixon and A. Sen, \nup292 (1987) 109;\nextline
J.A. Casas, E. Katehou and C. Mu\~noz, \nup317 (1989) 171.}}

\def\MOOR{\rrr\MOOR{  G. Moore and P. Nelson,
\prl53 (1984) 1519;
L. Alvarez-Gaum\'e and P. Ginsparg, \nup262 (1985) 439;
J. Bagger, D. Nemeschansky and S. Yankielowicz, \nup262 (1985) 478;
A. Manohar, G. Moore and P. Nelson, \plb152 (1985) 68;
W. Buchm\"uller and W. Lerche, Ann. Phys. {\bf 175} (1987) 159.}}

\def\MANOM{\rrr\MANOM{
W. Lerche, B. Nilsson and A. N. Schellekens, \nup299 (1988) 91;\nextline
J. A. Casas, E. Katehou and C. Mu\~ noz, \nup317 (1989) 171.}}
\def\FAYIL{\rrr\FAYIL{M. Dine, N. Seiberg and E. Witten, \nup289 (1987)
585;\nextline
J. Atick, L. Dixon and A. Sen, \nup292 (1987) 109;\nextline
M. Dine, I. Ichinose and N. Seiberg, \nup293 (1987) 253.}}
\def\REAL{\rrr\REAL{A. Font, L. E. Ib\' a\~ nez, H. P. Nilles and
F. Quevedo, \plb210 (1988) 101; \plb213 (1988) 564;\nextline
J. A. Casas and C. Mu\~ noz, \plb209 (1988) 214; \plb214 (1988) 63.}}
\def\FGN{\rrr\FGN{H. P. Nilles, \plb115 (1982) 193;\nextline
S. Ferrara, L. Girardello and H. P. Nilles, \plb125 (1983) 457.}}
\def\IN{\rrr\IN{L. E. Ib\' a\~ nez and H. P. Nilles, \plb169 (1986) 354;
\nextline
T. Taylor and G. Veneziano, \plb212 (1988) 147;
\nextline
L. Dixon, V. Kaplunovsky and J. Louis, SLAC-PUB-5138 (1990).}}
\def\SEMI{\rrr\SEMI{A. Font, L. E. Ib\' a\~ nez, D. L\" ust and
F. Quevedo, \plb245 (1990) 401;\nextline
S. Ferrara, N. Magnoli, T. Taylor and G. Veneziano, \plb245 (1990) 409.}}
\def\OLEC{\rrr\OLEC{H. P. Nilles and M. Olechowski, \plb248 (1990)
268;\nextline
P. Binetruy and M. K. Gaillard, preprint  CERN-TH.
5727/90.}}
\def\DGSB{\rrr\DGSB{L.E. Ib\'a\~nez and G.G. Ross, \nup368 (1992) 3.}}
\def\ILB{\rrr\ILB{L.E. Ib\'a\~nez and D. L\"ust, CERN-TH.6380 (1992).}}
\def\REFIN{\rrr\REFIN{U. Ellwanger, \nup238 (1984) 665 ;\nextline
H.J.Kappen, Phys.Rev.D38 (1988) 721;\nextline
G. Gamberini, G. Ridolfi and F. Zwirner, \nup331 (1990) 331.}}
\def\GIRAR{\rrr\GIRAR{L. Girardello and M. Grisaru, \nup194 (1982) 65.}}
\def\DREEB{\rrr\DREEB{M. Drees and M.M. Nojiri, KEK preprint
KEK-TH-290 (1991).}}
\def\BAHALL{\rrr\BAHALL{R. Barbieri and L. Hall, LBL preprint
LBL-31238 (1991).}}
\def\RR{\rrr\RR{R.G. Roberts and G.G. Ross, Rutherford Lab. preprint
RAL-92-005 (1991).}}
\def\DA{\rrr\DA{M. Daniel and J.A. Pe\~narrocha, \plb127 (1983) 219.}}

\def\PLANCK{\rrr\PLANCK{L.E. Ib\'a\~nez, \plb126 (1983) 196;
\nextline J.E. Bjorkman and D.R.T. Jones, \nup259 (1985) 533.}}

\def\ZNZM{\rrr\ZNZM{A. Font, L.E. Ib\'a\~nez and F. Quevedo,
\plb217 (1989) 272.}}

\def\IN{\rrr\IN{L.E. Ib\'a\~nez and H.P. Nilles, \plb169 (1986) 354.}}

\def\EINHJON{\rrr\EINHJON{M. Einhorn and D.R.T. Jones, \nup196 (1982)
                      475.}}

\def\KAPLU{\rrr\KAPLU{V. Kaplunovsky, \nup307 (1988) 145.}}

\def\AMALDI{\rrr\AMALDI{
J. Ellis, S. Kelley and D.V. Nanopoulos, \plb249 (1990) 441;
\plb260 (1991) 131; \nextline
P. Langacker,      ``Precision tests of the standard model''
Pennsylvania preprint UPR-0435T, (1990);\nextline
U. Amaldi, W. de Boer and H. F\"urstenau, \plt260 (1991) 447;\nextline
P. Langacker and M. Luo, Phys.Rev.D44 (1991) 817;\nextline
R.G. Roberts and G.G. Ross, talk presented by G.G. Ross at 1991 Joint
International Lepton-Photon Symposium and EPS Conference,     to
be published.
}}

\def\DG{\rrr\DG{S. Dimopoulos, S. Raby and F. Wilczek, Phys. Rev.
D24 (1981) 1681;\nextline
                 L.E. Ib\'a\~nez and G.G. Ross, \plb105 (1981) 439;
\nextline S. Dimopoulos and H. Georgi, \nup193 (1981) 375.}}

\def\IMNQ{\rrr\IMNQ{L.E. Ib\'a\~nez, H.P. Nilles and F. Quevedo,
\plt187 (1987) 25; L.E. Ib\'a\~nez, J. Mas, H.P. Nilles and
F. Quevedo, \nup301 (1988) 157; A. Font, L.E. Ib\'a\~nez,
F. Quevedo and A. Sierra, \nup331 (1990) 421.}}

\def\SCHELL{\rrr\SCHELL{A.N. Schellekens, \plt237 (1990) 363.}}

\def\GQW{\rrr\GQW{H. Georgi, H.R. Quinn and S. Weinberg, Phys. Rev.
Lett. ${\underline{33}}$ (1974) 451.}}

\def\GINS{\rrr\GINS{P. Ginsparg, \plt197 (1987) 139.}}

\def\ELLISETAL{\rrr\ELLISETAL{I. Antoniadis, J. Ellis, R. Lacaze
and D.V. Nanopoulos, {\it ``String Threshold Corrections and
                Flipped $SU(5)$'',} preprint CERN-TH.6136/91 (1991);
    S. Kalara, J.L. Lopez and D.V. Nanopoulos, {\it``Threshold
   Corrections and Modular Invariance in Free Fermionic Strings'',}
      preprint CERN-TH-6168/91 (1991).}}

\def\LOUIS{\rrr\LOUIS{J. Louis, {\it
         ``Non-harmonic gauge coupling constants in supersymmetry
         and superstring theory'',} preprint SLAC-PUB-5527 (1991);
         V. Kaplunovsky and J. Louis, as quoted in J. Louis,
         SLAC-PUB-5527 (1991).}}

\def\DIN{\rrr\DIN{J.P. Derendinger, L.E. Ib\'a\~nez and H.P Nilles,
        \nup267 (1986) 365.}}

\def\DHVW{\rrr\DHVW{L. Dixon, J. Harvey, C.~Vafa and E.~Witten,
         \nup261 (1985) 651;
        \nup274 (1986) 285.}}

\def\DKLB{\rrr\DKLB{L. Dixon, V. Kaplunovsky and J. Louis,
         \nup355 (1991) 649.}}

\def\DKLA{\rrr\DKLA{L. Dixon, V. Kaplunovsky and J. Louis, \nup329 (1990)
            27.}}

\def\ALOS{\rrr\ALOS{E. Alvarez and M.A.R. Osorio, \prv40 (1989) 1150.}}

\def\FILQ{\rrr\FILQ{A. Font, L.E. Ib\'a\~nez, D. L\"ust and F. Quevedo,
           \plt245 (1990) 401.}}

\def\CFILQ{\rrr\CFILQ{M. Cvetic, A. Font, L.E.
           Ib\'a\~nez, D. L\"ust and F. Quevedo, \nup361 (1991) 194.}}

\def\FILQ{\rrr\FILQ{A. Font, L.E. Ib\'a\~nez, D. L\"ust and F. Quevedo,
           \plt245 (1990) 401.}}

\def\DUAGAU{\rrr\DUAGAU{S. Ferrara, N. Magnoli, T.R. Taylor and
           G. Veneziano, \plt245 (1990) 409;\nextline  H.P. Nilles and M.
           Olechowski, \plt248 (1990) 268;\nextline  P. Binetruy and M.K.
           Gaillard, \plt253 (1991) 119;\nextline J. Louis, SLAC-PUB-5645
           (1991);\nextline S. Kalara, J. Lopez and D. Nanopoulos,
           Texas A\&M  preprint CTP-TAMU-69/91.}}

\def\CFILQ{\rrr\CFILQ{M. Cvetic, A. Font, L.E.
           Ib\'a\~nez, D. L\"ust and F. Quevedo, \nup361 (1991) 194.}}

\def\FIQ{\rrr\FIQ{A. Font, L.E. Ib\'a\~nez and F. Quevedo,
        \plt217 (1989) 272.}}

\def\FLST{\rrr\FLST{S. Ferrara,
         D. L\"ust, A. Shapere and S. Theisen, \plt225 (1989) 363.}}

\def\FLT{\rrr\FLT
{S. Ferrara, D. L\"ust and S. Theisen, \plt233 (1989) 147.}}

\def\IBLU{\rrr\IBLU{L. Ib\'a\~nez and D. L\"ust,
          \plb267 (1991) 51.}}

\def\GAUGINO{\rrr\GAUGINO{J.P. Derendinger, L.E. Ib\'a\~nez and H.P. Nilles,
            \plb155 (1985) 65;
         M. Dine, R. Rohm, N. Seiberg and E. Witten, \plb156 (1985) 55.}}

\def\GHMR{\rrr\GHMR{D.J. Gross, J.A. Harvey, E. Martinec and R. Rohm,
         \prl54 (1985) 502; \nup256 (1985) 253; \nup267 (1986) 75.}}

\def\ANT{\rrr\ANT{I. Antoniadis, K.S. Narain and T.R. Taylor,
        \plb267 (1991) 37.}}

\def\WITTEF{\rrr\WITTEF{E. Witten, \plb155 (1985) 151.}}

\def\HB{\rrr\HB{L. Hall and R. Barbieri, private communication and
preprint in preparation (1991).}}

\def\SCHELL{\rrr\SCHELL{A.N. Schellekens, ``Superstring
construction'', North Holland, Amsterdam (1989).}}

\def\FLIP{\rrr\FLIP{I. Antoniadis, J. Ellis, J. Hagelin and
D.V. Nanopoulos, \plb231 (1989) 65 and references therein. For a
recent review see J. Lopez and D.V. Nanopoulos,
                  Texas A\& M preprint CTP-TAMU-76/91 (1991).}}

\def\GKMR{\rrr\GKPM{B. Greene, K. Kirklin, P. Miron and G.G. Ross,
\nup292 (1987) 606.}}

\def\OTH{\rrr\OTH{D. Bailin, A. Love and S. Thomas, \plb188 (1987)
193; \plb194 (1987) 385; B. Nilsson, P. Roberts and P. Salomonson,
\plb222 (1989) 35;
J.A. Casas, E.K. Katehou and C. Mu\~noz, \nup317 (1989) 171;
J.A. Casas and C. Mu\~noz, \plb209 (1988) 214, \plb212 (1988) 343
J.A. Casas, F. Gomez and C. Mu\~noz, \plb251 (1990) 99;
A. Chamseddine and J.P. Derendinger, \nup301 (1988) 381;
A. Chamseddine and M. Quiros, \plb212 (1988) 343, \nup316 (1989) 101;
T. Burwick, R. Kaiser and H. M\"uller, \nup355 (1991) 689;
Y. Katsuki, Y. Kawamura, T. Kobayashi, N. Ohtsubo,
Y. Ono and K. Tanioka, \nup341 (1990) 611.}}

\def\SCHLUS{\rrr\SCHLUS{For a review, see e.g. J. Schwarz, Caltech
preprint CALT-68-1740 (1991); D. L\"ust,  CERN preprint TH.6143/91.
}}

\def\LMN{\rrr\LMN{J. Lauer, J. Mas and H.P. Nilles, \plb226 (1989)
251, \nup351 (1991) 353; W. Lerche, D. L\"ust and N.P. Warner,
\plb231 (1989) 417.}}

\def\ILR{\rrr\ILR{ L.E. Ib\'a\~nez, D. L\"ust and G.G. Ross,
\plb272 (1991) 251.}}

\def\ANTON{\rrr\ANTON{I. Antoniadis, J. Ellis, S. Kelley and
D.V. Nanopoulos, \plb272 (1991) 31 .}}

\def\HIGH{\rrr\HIGH{D. Lewellen, \nup337 (1990) 61;
J.A. Schwartz, Phys.Rev. D42 (1990) 1777.}}

\def\HIGHK{\rrr\HIGHK{A. Font, L.E. Ib\'a\~nez and F. Quevedo,
\nup345 (1990) 389; J. Ellis, J. Lopez and D.V. Nanopoulos,
\plb245 (1990) 375.}}

\def\LLR{\rrr\LLR{C.H. Llewellyn-Smith, G.G. Ross and
J.F. Wheater, \nup177 (1981) 263; \nextline
S. Weinberg, \plb91 (1980) 51; \nextline
L. Hall, \nup178 (1981) 75;\nextline
P. Binetruy and T. Schucker, \nup178 (1981) 293.}}

\def\IL{\rrr\IL{  L.E. Ib\'a\~nez and D. L\"ust, CERN-TH.6380/92
(1992).}}

\def\YO{\rrr\YO{ L.E. Ib\'a\~nez, {\it``Some topics
in the low energy physics from superstrings''} in proceedings of
the NATO workshop on ``Superfield Theories", Vancouver,
Canada. Plenum Press, New York (1987).}}

\def\SDUAL{\rrr\SDUAL{A. Font, L.E. Ib\'a\~nez, D. L\"ust
and F. Quevedo, \plb249 (1990) 35.}}

\def\MALLOR{\rrr\MALLOR{For a recent review see L.E. Ib\'a\~nez,
{\it ``Beyond the Standard Model (yet again)''}, CERN preprint
TH.5982/91, to appear in the Proceedings of the 1990 CERN
School of Physics, Mallorca (1990).}}

\def\FIQS{\rrr\FIQS{A. Font, L.E. Ib\'a\~nez, F. Quevedo and
A. Sierra, \nup337 (1990) 119.}}

\def\LT{\rrr\LT{D. L\"ust and T.R. Taylor, \plb253 (1991) 335;
B. Carlos, J. Casas and C. Mu\~noz, preprint CERN-TH.6049/91
(1991).}}

\def\DGSA{\rrr\DGSA{L.E. Ib\'a\~nez and G.G. Ross, \plb260 (1991) 291.}}

\def\HLW{\rrr\HLW{L. Hall, J. Lykken and S. Weinberg, Phys.Rev.D27 (1983)
  2359.}}

\def\MBMT{\rrr\MBMT{ G. Lazarides and Q. Shafi, Bartol Research
preprint BA-91-25 (1991);\nextline
S. Kelley, J. L\'opez and D. Nanopoulos, Texas AM preprint
CTP-TAMU-79-91 (1991);\nextline
H. Aranson, D. Casta\~no, B. Keszthelyi, S. Mikaelian, E. Piard,
P. Ramond and B. Wright, Phys.Rev.Lett. 67 (1991) 2933;\nextline
S. Dimopoulos, L. Hall and S. Raby, LBL-31441 UCB-PTH 91/61.}}

\def\SUSYGUT{\rrr\SUSYGUT{E. Witten, \nup188 (1981) 513;\nextline
S. Dimopoulos and H. Georgi, \nup193 (1981) 150;\nextline
N. Sakai, Z.Phys. C11 (1982) 153;\nextline
E. Witten, \plb105 (1981) 267;\nextline
L.E. Ib\'a\~nez and G.G. Ross, \plb105 (1981) 439; \plb110 (1982) 215
\nextline
L. Alvarez-Gaum\'e, M. Claudson and M. Wise, \nup207 (1982) 16;\nextline
M. Dine and W. Fischler, \nup204 (1982) 346;\nextline
J. Ellis, L.E. Ib\'a\~nez and G.G. Ross, \plb113 (1982) 283;\nup221
(1983)  29;\nextline
C. Nappi and B. Ovrut, \plb113 (1982) 175;\nextline
S. Dimopoulos and S. Raby, \nup219 (1983) 479;\nextline
J. Polchinski and L. Susskind, Phys.Rev.D26 (1982) 3661;\nextline
J. Ellis, D. Nanopoulos and K. Tamvakis \plb121 (1983) 123;\nextline
H.P. Nilles, \nup217 (1983) 366.}}

\def\AAA{\rrr\AAA{J.M. Frere, D.R.T. Jones and S. Raby, \nup222 (1983)
11.}}

\def\AAB{\rrr\AAB{M. Drees, M. Gl\"uck and K. Grassie, \plb157 (1985)
164.}}

\catcode`@=12
\newtoks\Pubnumtwo
\newtoks\Pubnumthree
\catcode`@=11
\def\p@bblock{\begingroup\tabskip=\hsize minus\hsize
   \baselineskip=0.5\ht\strutbox\topspace-2\baselineskip
   \halign to \hsize{\strut ##\hfil\tabskip=0pt\crcr
   \the\Pubnum\cr  \the\Pubnumtwo\cr %\the \Pubnumthree\cr
%  \the\date\cr
   \the\pubtype\cr}\endgroup}
\pubnum={6501/92}
\date{May   1992   }
\pubtype={}
\titlepage
\vskip -.6truein
\title{ Computing the Weak Mixing Angle from Anomaly Cancellation}
 %\vskip -.5truein
 \vskip 0.05truein
 \centerline{\bf Luis E. Ib\'a\~nez}
 \vskip .1truein
 \centerline{CERN, 1211 Geneva 23, Switzerland}
 \vskip .1truein
\abstract\noindent\nobreak

I remark that the weak mixing angle  in the standard model may be
computed even in the absence of a grand unification symmetry.
In particular, if there is an additional gauged $U(1)$ symmetry at some
large scale which can be made anomaly-free only by a Green-Schwarz (GS)
mechanism, this typically results in a prediction for the weak angle.
In the case of the  standard model one can see that the standard
Peccei-Quinn symmetry may be gauged and the anomalies cancelled through
a GS mechanism. Remarkably enough, cancelation of anomalies works only
for the `canonical' value $sin^2\theta _W=3/8$. In the case of the
supersymmetric standard model one can also find $U(1)$ currents which
may be made anomaly-free through a GS mechanism but the canonical value
is only obtained in the absence of any Higgs multiplet.
If the analysis is extended to include $U(1)$ R-symmetries, there
is a unique class of $U(1)$s which give rise to the canonical
value. The R-symmetry is only anomaly-free for
$sin^2\theta _W=(4N_g-3)/(10N_g-3N_D-3)$, where $N_g,N_D$ are
the number of generations and Higgs pairs. The natural context
in which the above scenario may naturally arise is string theory.
I also emphasize other interesting possibilities offered by the
GS mechanism to model-building.

\pagenumber=1
\sequentialequations

The most interesting predictions coming out from standard grand unified
theories are charge quantization and the prediction for the
weak mixing angle $sin^2\theta _W=3/8$ \GQW . More recently \VIRGIN\
it has been realized that charge quantization (the fact that quarks have
charges $\pm 1/3,\pm 2/3$ in units of the electron charge) may be
obtained as a direct consequence of the cancellation of all
standard model (SM) anomalies, including the gravitational ones
\GRA . On the other hand, the `canonical' prediction for the weak angle
seems to be unavoidably tied up to the embedding of the
SM into a standard GUT of the type of $SU(5), SO(10)$ or
$E_6$. It is an interesting question whether    this is indeed
the case or there are alternative situations in which
a definite prediction for the weak angle may be obtained, even
without any grand unification symmetry. After all,   the standard
GUT models, even in their supersymmetric versions, have some
difficult problems for which no convincing solution has been
given since they were first formulated. One of them is the
notorious doublet-triplet splitting of the Higgs multiplet
and another the quark-lepton mass relationships of the first
generation. While the latter may be solved by apropriately
complicating the Higgs sector, this looks more like adding
an `epicycle' than a real solution. At this point one might be
tempted to disposse alltogether  with the unification idea but
the amazing agreement of the $sin^2\theta _W$ prediction with
recent LEP data (at least for the supersymmetric case) makes
one to hesitate before taking such a radical attitude.
In this situation is clearly important to search for alternative
mechanisms to predict the correct weak angle without having
to buy the complete grand unification idea and its problems.

In the present note I discuss a        mechanism in which
the value of the mixing angle is determined by the anomaly
structure of the theory. In particular, if there is an
additional gauged $U(1)$ symmetry at some large scale which can
only be made anomaly-free by a Green-Schwarz (GS) mechanism \ANOM ,
this typically
results in a prediction for the weak mixing angle. This extra
$U(1)$ symmetry is generically spontaneously broken at a large
scale although it may survive as a global symmetry.
The simplest example of this mechanism is provided by gauging the
Peccei-Quinn symmetry in the two-Higgs standard model (SM).
This simplest extension of the SM can be made anomaly-free
through a GS mechanism only if the canonical $sin^2\theta _W=3/8$
is taken.

          In the case of the minimal supersymmetric standard model (SSM),
one can also find $U(1)$ gauge symmetries which can be made anomaly-free
     through a GS mechanism but the canonical $sin^2\theta _W=3/8$
is only obtained for the unrealistic case of no Higgs doublet present.
However,     in the supersymmetric case the gauginos may also be
charged with respect to a $U(1)$ symmetry if one allows for
R-symmetries. Such R-symmetries cannot be gauged in the usual sense,
since that would explicitely break SUSY, but may be gauged as
sigma-model symmetries, as is generically the case in
$N=1$ supergravity lagrangians. These $U(1)$ R-symmetries have
in general mixed anomalies with respect to the SM gauge interactions.
Requiring       the cancellation of these mixed anomalies through
a GS mechanism again leads to   predictions for the weak angle.
There is a unique class of $U(1)$ R-symmetries which gives rise
to the canonical result for $sin^2\theta _W$. It is such that
i)  only the minimal pair of Higgs multiplets (and no more) is
compatible with anomaly cancellation and ii) all B and/or L-violating
Yukawa couplings are forbidden. Finally, I also discuss some
other possible applications of the four-dimensional GS mechanism
to model-building.

A crucial role in the propossed mechanism is played by the
Green-Schwarz anomaly cancellation mechanism \ANOM . This mechanism
allows for gauged $U(1)$ currents whose anomalies, as naively
computed through the triangle graphs, are non-vanishing. The
anomalies are in fact cancelled by assigning a non-trivial
gauge transformation to an axion $\eta (x) $ introduced in the theory
which couples universally to all gauge groups \FDGS . The quadratic
gauge piece of the Lagrangian has the form
$$
{1\over {g^2(M)}}\sum _{i=1,2,3,X} k_i\ F^2_i\ +\
i\  {\eta (x) }        \ \sum _{i=1,2,3,X} k_i\ F_i{\tilde F}_i \ ,
\eqn\gaug
$$
where $g$ is the gauge coupling constant at some (presumably large)
scale $M$, $F_i$ are the gauge field strengths
and $k_i$ are in principle real numbers which take into account
the different normalization of the gauge group generators.
Notice that the action in eq.\gaug \ applies only at some  unknown
large scale $M$ at which the coupling constants of the
different gauge groups are assumed to meet (up to the $k_i$
normalization factors). Indeed, this assumption (as well as
the existence of a unique axion) is crucial in order to relate
the cancellation of anomalies to the normalization of
coupling constants at the given scale $M$.
                    However, it turns out to be a natural situation in
4-D strings. Notice that, on the other hand, the above action does
not assume   ` a priori' any GUT-like symmetry relating the
different $k_i$s.
                 Below the $M$     scale, the coupling constants will
run as usual according to their renormalization group equations.
            In the case of string theory this scale $M$ is
related to the Planck mass but we will not assume any definite
 value here, but simply assume that it exists and it is large.
The index
$i$ runs over the three gauge groups $U(1)\otimes SU(2)\otimes
SU(3)$ of the SM and the extra `anomalous' gauge group $U(1)_X$.
Under a $U(1)_X$ gauge transformation one has
$$
\eqalign{
A_X^{\mu }\ &\rightarrow \ A_X^{\mu }\ +\ \partial ^{\mu }\theta (x) \cr
\eta \ &\rightarrow \   \eta \ -\ \theta (x) \delta _{GS} \cr }
\eqn\gtrans
$$
where $\delta _{GS}$ is a constant and $\eta (x)$ is the axion field.
If the coefficients $C_i$ of the mixed $U(1)_X$-$SU(3)$,-$SU(2)$,-$U(1)$
are in the ratio
$$
{{C_1}\over {k_1}}\ =\ {{C_2}\over {k_2}}\ =\ {{C_3}\over {k_3}}\
=\ \delta _{GS} \ ,
\eqn\cond
$$
those mixed anomalies will be cancelled by the gauge variation of
the second term in eq.\gaug . Since there may be in the spectrum
extra singlet particles with $U(1)_X$ quantum numbers but no
SM gauge interactions, we will not consider here the equivalent
conditions involving the $U(1)_X$ anomaly coefficient, since those
singlets can always be chosen so that that anomaly is cancelled.
For the same reason we will not consider the
                                mixed $U(1)_X$-gravitational
anomalies.  On the other hand, to be consistent, one has to impose
that the mixed $U(1)_Y-U(1)_X^2$ anomaly vanishes identically since
it only involves standard model fermions and cannot be cancelled by
a GS mechanism.

           The anomaly cancellation mechanism sketched above
appears most naturally in the context of four-dimensional
superstrings. There there is always a unique axion field
$\eta (x)$ with the couplings in eq.\gaug  and there is also
a scalar dilaton field $\phi $ with a vev. $<\phi >=1/g^2$ coupling to
$F_i^2$.
The coefficients $k_i$ are the Kac-Moody levels of the corresponding
gauge algebra \GINS . For the case of non-abelian groups like
$SU(3)$ and $SU(2)$ those levels are integer and in practically
all models constructed up to now one has $k_2=k_3=1$. In the
case of an abelian group like $U(1)$-hypercharge, $k_1$ is a
normalization factor (not necessarily integer) and is model
dependent. In string theories the presence of $U(1)$s whose
anomalies are cancelled through a GS mechanism is also ubiquitous
(see e.g. ref.\IMNQ ).
        For the reader familiar with string construction,
           it is perhaps worth showing an specific 4-D string with
an `anomalous' $U(1)$ in which relationships analogous to those
in eq.\cond \  are obtained. Consider as an example one of the three
$(0,2)$ embeddings of the $Z_3$ orbifold, the one leading to the
gauge group $E_7\times SO(14)\times U(1)_1\times U(1)_A$. This may
be obtained by acting on the $E_8\times E_8$ degrees of freedom with a
shift ${\vec v}=1/3(1,1,0,0,...,0)\times (2,0,0,...,0)$. The charged
chiral spectrum contains 3 copies of chiral multiplets with quantum
numbers $(56,1,0,1)+(1,1,0,-2)+(1,14,-2,0)+(1,64,1,0)$ from
the untwisted sector; 27 copies of $(1,14,-2/3,2/3)+(1,1,4/3,-4/3)$
from the twisted sector and 81 copies of $(1,1,4/3,2/3)$ from
twisted oscilator states \IMNQ . One can easily check that all gauge
interactions are anomaly-free except for $U(1)_A$ which is
anomalous. The mixed anomalies of $U(1)_A$ with $E_7,SO(14)$ and
$U(1)_1$ yield respectively $C_7=18$, $C_{14}=18$ and $C_1=144$.
Since this anomaly has to be cancelled through a GS mechanism,
one necessarily has, in analogy with eq.\cond \ ,
$k_1=C_1/C_7=C_1/C_{14}=8$, if anomalies are to be cancelled
(recall that in this type of compactifications the Kac-Moody level
of the non-Abelian groups is equal to one).
Indeed, this indirect way of computing $k_1$ is consistent
with the direct computation (see e.g. Appendix C in ref.\FENO\ )
                            from gauge coupling string unification \GINS
. It is easy to show that the $U(1)_1$ generator may be written in terms
of the 16 Cartan subalgebra coordinates of $E_8\times E_8$ as
$Q_1=i\sum _I Q_I\partial X_I $ with
${\vec {Q } }=(0,0,....,0)\times (2,0,...,0)$. From here one finds
$k_1=2\sum _IQ_I^2=8$, as it should. In order to compute $k_1$
from the first method one only needs to know the quantum numbers
of the $massless$ spectrum  whereas the second requires the
knowledge of the specific vertex operator associated to
$U(1)_1$.
We emphasize again that, although string theory is the natural
scheme in which the $GS$ mechanism naturally appears,
                    it is not extrictly necessary to assume
that an underlying string theory exists and it is not necessary
either to assume $N=1$ supersymmetry. The idea we want to stress in
this note is that there exist simple Lagrangians  $not$ involving
a unification group like $SU(5)$ in which consistency constrains
the possible values of the tree-level weak angle.

{}From eqs.\gaug\ and \cond\ one obtains for the tree level
weak angle at the large scale $M$
$$
sin^2\theta _W\ =\ {{k_2}\over {k_1+k_2}}\ =\
{{C_2}\over {C_1+C_2}} \ .
\eqn\wein
$$
The above expression shows that, for each given `anomalous' $U(1)$,
the cancellation of the anomalies through a GS mechanism gives
a definite prediction for the weak angle in terms of the
coefficients of the anomaly. The latter may be computed in terms
of the $U(1)_X$ charges of the $massless$ fermions of the theory.

The next obvious question is whether there is any global
$U(1)$ symmetry in the SM (or in the SSM) with      an anomaly
which can potentially be cancelled by a GS mechanism. To see that
let us consider the most general global symmetry of the SM. It
can be generated by a linear  combination of $U(1)$ charges
(up to  linear combinations with hypercharge)
$$ Q_X\ =\ mR\ +   \   nA \ + \ \sum _{i=1}^{N_g} p_iL_i
\eqn\qe
$$
where $m,n$ and $p_i$ are real numbers and $N_g$ is the number of
generations. The charges of the standard model fermions with respect to
$R,A$ and $L_i$ are shown in table 1 (the notation is analogous to that
in ref.\DGSB ). The generator $R$ corresponds
to the third component of a right-handed isospin whereas
the $L_i$ are the usual lepton numbers. The generator $A$
corresponds to the standard Peccei-Quinn symmetry and can only be
a symmetry in the two-Higgs version of the SM (otherwise the u-quarks
would remain massless). Notice that the baryon number symmetry
may be writen in terms of $R,L_i$ and the weak hypercharge.
The standard model has three $U(1)$ symmetries with no
mixed anomalies with the $SU(3)\otimes SU(2)\otimes U(1)$ interactions.
Those are $B-L$ (or, equivalently, up to a hypercharge rotation,
                                   $R$) and two linear combinations
of the three lepton numbers (e.g. $L_{\mu }-L_{\tau }$ ).
In fact the mixed anomalies of the
general $Q_X$ symmetry are given by
$$
\eqalign{
C_3\ &=\ -n {{N_g}\over 2} \cr
C_2\ &=\ -n {{N_g}\over 2}\ -\ {{\sum _i^{N_g} p_i}\over 2} \cr
C_1\ &=\ -n {5\over 6}N_g\ +\ {{\sum _i^{N_g} p_i}\over 2}    \cr }
\eqn\ansm
$$
where $N_g$ is the number of generations.
Impossing $C_3/k_3=C_2/k_2$ one gets the constraint
$$
\sum _i^{N_g} p_i\ =\ nN_g\ ({{k_2}\over {k_3}}  - 1      )\ .
\eqn\kdkt
$$
There is an extra constraint  coming from impossing      cancellation
of the mixed $U(1)_Y-U(1)_X^2$ anomalies, which have to vanish
identically. This yields the condition $\sum p_i=mN_g/(m-1)$ which
combined with eq.\kdkt \ yields
$$
{{k_2}\over {k_3}}\ =\ 1 \ +\ {{m}\over {m-1}} \ .
\eqn\any
$$
Finally, impossing $C_2/k_2= C_1/k_1$ one gets
$$
{{k_1}\over {k_2}}\ =\ {{5/3\ -\ {m\over {m-1}} }\over
{1\ +\ {m\over {m-1}}}}
\eqn\kukd
$$

The simplest $U(1)$ symmetry whose mixed anomalies may be cancelled by
a GS mechanism corresponds to the Peccei-Quinn generator $A$, i.e. $m=0$.
In this case $\sum p_i=0$ and anomaly cancellation predicts:
$$
\eqalign{
 {{k_2}\over {k_3}}\ &=\ 1  \cr
{{k_1}\over {k_2}}\ &=\ {5\over 3} \ . \cr }
\eqn\cucu
$$
This result gives rise to the canonical prediction $sin^2\theta _W=3/8$.
The fact that asking for anomaly cancellation with respect to the
simple gauged charge $A$ leads to this canonical result is quite
remarkable.

What is the fate of the extra $U(1)_X$ interaction? The structure
of the GS mechanism forces this gauge boson to become massive by
swallowing the axion field as its longitudinal component \FDGS .
This is more clearly seen in the dual formulation of the axion field in
terms of a two index antisymmetric tensor $B_{\mu \nu }$. The field
strength of this tensor $H_{\mu \nu \rho }$ (which contains the
standard gauge Chern-Simons  term \FDGS )
is related to the axion
field by $\partial _{\mu }\eta (x)=      \epsilon _{\mu \nu \rho \sigma }
H^{\nu \rho \sigma }$. In this equivalent formulation    the anomaly
cancellation mechanism requires a one-loop counterterm in the
Lagrangian of the form $M^2\epsilon _{\mu \nu \rho \sigma }B^{\mu \nu }
F^{\rho \sigma }$. After the duality transformation this term
becomes $M^2\partial _{\mu } \eta A_X^{\mu }$ in terms of the axion.
This is nothing but a typical Higgs mechanism term which gives a
mass $M$ to the gauge boson $A_X$. In string theory, the role of radial
mode in the Higgs mechanism is played by a dilaton field, which, as
discussed above, is always present in the supersymmetric case.
Notice, however that the
$U(1)_X$ symmetry may remain  as an effective $global$ symmetry in
the low energy Lagrangian. That depends on wether there are
or not additional singlet scalar fields with X-charge and non-vanishing
vevs in the theory.
                                        In the case of the relevant
$A$ generator discussed above, the residual global symmetry would
guarantee
the absence of flavour changing Higgs-mediated neutral currents. Indeed
this is so because the $A$ symmetry forces that the $u$-quarks
get their masses from a different Higgs field than the $d$-quarks.
This is an interesting byproduct of the symmetry.
Notice also that for the above mechanism to work one needs to
have two Higgs doublets in the SM, a prediction which should be
experimentally testable.

Due to the notorius `gauge hierarchy' problem one might prefer
a supersymmetric version of the standard model. The previous
analysis is straightforwardly extended to the SSM. The only new
ingredient is the existence \FDGS\ of a one-loop Fayet-Iliopoulos term
(SUSY-counterpart of the $B_{\mu \nu }F_{\rho \sigma }$ coupling)
which plays no role in our discussion. In the SUSY case
new fermions may contribute to the mixed anomalies, the Higgsinos
(the gauginos only contribute to R-symmetry anomalies, see below).
The most general $U(1)_X$ generator coupling to the
SSM is still given by the expression in eq.\qe . Although the
addition of the Higgsino fields could allow for new generators,
we fix the charge assignements of those fermions by impossing that
the standard Yukawa couplings which give masses to quarks and
leptons are allowed \DGSB . The coefficients of the mixed anomalies are
now $$
\eqalign{
C_3\ &=\ -n{{N_g}\over 2}   \cr
C_2\ &=\ -n{{N_g}\over 2}\ -\ {{\sum _i^{N_g}p_i}\over 2}\ +\
     n{{N_D}\over 2}  \cr
C_1\ &=\ -n{5\over 6}N_g\ +\ {{\sum _i^{N_g}p_i}\over 2}\
 +\  n{{N_D}\over 2}      \cr}
\eqn\anssm
$$
where $N_D$ is the number of Higgs pairs.
Requiring the vanishing    of the mixed $U(1)_Y-U(1)_X^2$ anomalies
gives the constraint (without loss of generality we set $n=1$) :
$$
\sum _{i=1}^{N_g} p_i\ =\ {{N_D}\over {2(m-1)}}\ +\
{m\over {m-1}}(N_g-N_D)  \ ,
\eqn\anyy
$$
and further impossing $C_2/k_2=C_3/k_3$ yields the result
$$
{{k_2}\over {k_3}}\ =\ 1\ +\ {{N_D}\over {N_g}}{{3-4m}\over {2(m-1)}}\ +\
{m\over {m-1}}  \ .
\eqn\coci
$$
Finally, impossing
$C_1/k_1=C_2/k_2$ one gets the result
$$
{{k_1}\over {k_2}}\ =\ {{5/3\ -\ {m\over {m-1}}\ +\ {{N_D}\over
{2N_g(m-1)}}  }\over {1\ +\ {{N_D}\over {N_g}} {{3-4m}\over {2(m-1)}}
\ +\  {m\over {m-1}}  } }
\eqn\kiki
$$
Notice that in the supersymmetric case the generator $A$ (i.e. $m=0$)
no longer gives the canonical result for the physical values
$N_g=3, N_D=1$, instead one finds $k_1/k_2=3$   and  $k_2/k_3=1/2$,
not very promising values. Indeed, if one insists in impossing
the string-like boundary condition    $k_2=k_3$ one finds
$m=-3/2$ and $k_1/k_2=1$, again not very interesting results.
Only in the unphysical no-Higgs case $N_D=0$ one can again gauge
the Peccei-Quinn current $A$ and obtain $sin^2\theta _W=3/8$.
Of course, one can also envisage other modifications of the
massless particle content of the SSM which may yield the
canonical (or other) values for the weak mixing angle.

The above result concerning the SSM looks a bit deceptive since it is in
fact for the supersymmetric model for which the canonical value of
the weak angle leads to better experimental agreement\ ! In fact,
in addition to the gauged $U(1)$ symmetries discussed above,
         supersymmetry allows for more general classes of
$U(1)$ symmetries.
In particular one can also consider assigning quantum numbers to the
gauginos, i.e. continuous $U(1)$ R-symmetries. The most general \DGSB\
such R-symmetry of the SSM may be writen as follows:
$$
Q_X\ =\ P\ \oplus \ mR\ \oplus \ nA\ \oplus \
\sum _{i=1}^{N_g}\ p_iL_i \ ,
\eqn\genss
$$
where $P$ is the R-symmetry generator defined in table 1. Notice
that the Higgsino charges are again fixed so us to allow for
the ordinary Yukawa couplings (the charges of the scalar SUSY partners
are increased in one unit  with respect to the corresponding fermion).
A $U(1)$ charge as in eq.\genss \  cannot be gauged in the usual
sense since it does not commute with supersymmetry (SUSY partners
have different charges). On the other hand this kind of
R-symmetries may be gauged in the `sigma-model sense'. This
means that the theory may posses an auxiliary gauge connection
which couples to this charge but does $not$ propagate, i.e.
there is no physical gauge boson asociated to it. This is in fact
the generic case in theories coupled to $N=1$ supergravity. The
kinetic terms of the scalar fields in those theories have a
sigma model structure and there is an auxiliary connection
associated to Kahler invariance and another independent connection
associated to the holonomy of the sigma model manifold.
These sigma model symmetries may have mixed anomalies \MOOR\ when
combined with the $SU(3)\times SU(2)\times U(1)$ interactions.
The   mixed anomalies may again     be cancelled through a GS
mechanism if an axion $\eta $  with couplings as in eq.\gaug \
transforms in the apropriate way under a sigma model gauge
transformation \DFKZ ,\IL .
This mechanism for cancelling the mixed sigma model anomalies
is very common in the context of four-dimensional strings. Thus,
for example, in all the thousands of models based on the $Z_3$ and
$Z_7$ orbifolds there  are mixed sigma model anomalies which are
always cancelled by a GS mechanism. This has been used in ref.\IL\
to show that the minimal SSM cannot be ever obtained from any of
those two large classes of orbifolds since the particle content
of the SSM cannot yield cancellation of mixed sigma model
anomalies.
Again, an example may be usefull for the reader familiar with
string model-building. Consider again the same $Z_3$ orbifold
model described above. One can see there is an auxiliary
conection $(T+T^*)^{-1}\partial ^{\mu }(T-T^*)$, in which
$T$ is the overall modulus complex field whose real part
gives the compactification radius of the orbifold.
This auxiliary field couples in the $N=1$ supergravity Lagrangian to
an $R$-symmetry which assignes quantum numbers as follows:
charge $-3$ for gauginos, $-1$ for twisted fermions, charge $+1$ for
untwisted fermions and charge $-3$ for twisted oscilators.
The mixed anomalies of this $R$-symmetry with $E_7,SO(14),U(1)_1$
and $U(1)_X$ yield respectively
$C_7=C_{14}=-36$, $C_1=-288$ and $C_X=-144$. Cancellation of sigma-model
anomalies through a Green-Schwarz mechanism requires
$k_1=C_1/C_7=C_1/C_{14}=8$, as it should.
The necessity of anomaly cancellation in this $Z_3$ orbifold
example is related to the necessary cancellation of
target space `duality anomalies' in orbifold compactifications.
We see how
impossing cancellation of mixed sigma model anomalies through a GS
mechanism gives us an alternative method to obtain a prediction
for the weak mixing angle in the supersymmetric standard
model. In this case one assumes that at some large scale
the scalar sector of the theory has a sigma model structure
which is made anomaly-free through a GS mechanism. The relevant
mixed anomalies are computed in exactly the same way as in the
gauge boson case but now we have the extra generator $P$ of
eq.\genss \ in the game. One finds:
$$
\eqalign{
C_3\ &=\ 3\ -\ 2N_g\ -n{{N_g}\over 2} \cr
C_2\ &=\ 2\ -\ 2N_g\ -n{{N_g}\over 2}\ +\ N_D(1+{n\over 2})
      \ -\ {{\sum _{i=1}^{N_g} p_i}\over 2}   \cr
C_1\ &=\ -{{10}\over 3}N_g \ -\ n{5\over 6}N_g\ +\ N_D(1+{n\over 2})
      \ +\ {{\sum _{i=1}^{N_g}p_i}\over 2}\ . \cr }
\eqn\anrss
$$
For the preferred $k_2=k_3$ case, the condition $C_2=C_3$ yields
$$
\sum _{i=1}^{N_g} p_i\ =\ 2(N_D-1)\ +\ nN_D
\eqn\condp
$$
and then one gets
$$
{{k_1}\over {k_2}}\ =\ {{C_1}\over {C_2}}\ =\
{{{5\over 6}(n+4)N_g+1-(n+2)N_D}\over {(2+{n\over 2})N_g-3}}  \ .
\eqn\kukdr
$$
For the physically interesting case $N_g=3, N_D=1$, the canonical
value $k_1/k_2=5/3$ is only obtained for $n=4$. Thus the R-symmetry
generator
$$
Q_{RX}\ =\ P\ \oplus \ 4(A\ \oplus L_i)
\eqn\qse
$$
where $L_i$ is any of the three lepton numbers,
can be made anomaly-free through a GS mechanism and yields the
canonical mixing angle for $N_g=3,N_D=1$. In particular such a
symmetry may be made anomaly-free provided
$$
sin^2\theta _W\ =\ {{4N_g\ -\ 3}\over {10N_g\ -3N_D\ -3}}
\eqn\sins
$$
which yields $sin^2\theta _W=3/8$ for $N_g=3,N_D=1$. Other
generators can be constructed which yield results different to the
canonical one but quite close to it, so that they can still yield
interesting results for the low energy couplings.
For example, for $n=\sum p_i=6$, one gets $k_2=k_3$ and $k_1/k_2=3/2$.
This leads, via the renormalization group, to the weak scale results
$sin^2\theta _W=0.235, \alpha _s=1/8$ and a unification scale
$M=0.12\times 10^{18}$GeV.
It is worth remarking that, in the absence of a GUT symmetry,
the value of $k_1$ need not be equal to the canonical value.
Perhaps other values like the one above could even be more interesting,
since they allow for unification scales closer to the Planck mass,
which is the natural scale in string models.

Notice that if we impose that the SSM should be invariant and
anomaly-free with respect to the sigma model current $Q_{RX}$
in eq.\qse ,
this can only be achieved  if there is just a single pair of
Higgs doublets, no more and no less. This is interesting because
this could provide a rationale for the existence of the minimal
set of Higgses, a mistery in all known unification scenarios.
A second interesting property of this symmetry is that it explicitly
forbids any possible dimension four term violating either
B or L-number. Thus     superpotential terms of the type
$udd$,$QdL$ or $LLe$ are necessarily absent. This is very interesting
because it is well known that in the SSM       the pressence of
some symmetry is needed anyway to avoid fast proton decay. In the above
scenario the symmetry which guarantees proton stability
automatically gives you the canonical weak mixing angle and
a rationale for the existence of the minimal set of Higgs doublets.

I  have addressed in this note the question whether the
succesfull GUT prediction for the weak mixing angle may be
obtained in other type of theories not involving a unification
group like $SU(5)$ or $SO(10)$. I  have shown that indeed this is
the case and that one can obtain definite predictions for the
weak mixing angle (including the canonical result $sin^2\theta _W=3/8$)
as a direct consequence of the cancellation of the anomalies in the
theory. This requires the existence of $U(1)$ symmetries at
some large scale whose mixed anomalies are cancelled through a
Green-Schwarz mechanism. Generically no remnant of the said symmetry will
remain at low energies, and its most relevant role will be to
fix the boundary conditions for the normalization of the
coupling constants. The mechanism requires some unification of
coupling constants (see eq.\gaug \ ) at some large unknown
scale $M$ but, $unlike$ GUTs, there is no large simple group like
$SU(5)$ fixing the relative normalizations ($k_i$) of coupling constants.
Ruther those normalizations are fixed by the requirement of
anomaly cancellation. The natural place where all these ingredients
appear is string theory in which the scale $M$ is expected to
be related to the Planck mass.

Let us end this letter with some comments on different possible
applications of the above ideas. One possible extension is to
`discrete gauge symmetries' \WK . Indeed, discrete gauge symmetries
should obey certain anomaly cancellation conditions \IR\ and they can
also be cancelled through a discrete version of the GS mechanism \DGSB .
Again, if the discrete gauge anomalies  are cancelled in that way,
constraints on the gauge couplings normalizations $k_i$  at the
scale $M$ are obtained. In this case, the anomaly cancellation conditions
are only verified modulo $N$ (for a $Z_N$ discrete gauge symmetry)
and the obtained conditions are weaker.

Independently of the application
to the computation   of the weak angle, I would like
to emphasize  the new possibilities offered by the GS mechanism
in model-building. Indeed, many $U(1)$ symmetries have been
discarded as possible gauged symmetries in the past because of
being anomalous. The four-dimensional version of the Green-Schwarz
mechanism allows to rescue some of those. They may be gauged
and become massive, as discussed above, at scale slightly
below $M$. In principle, nothing stops us from chosing e.g.
a scale $M\simeq O(1)$TeV, in which case the massive gauge
boson corresponding to the `anomalous' symmetry could be detected
experimentally.

                An specially interesting case is the standard
Peccei-Quinn symmetry of the two-doublet standard model, as we discussed
above. We have shown that this symmetry may be gauged and its
anomalies cancelled through a GS mechanism. This could have
interesting implications in the CP structure of the theory.
Notice that gauging the Peccei-Quinn symmetry makes it stable
against strong gravitational effects, which has been the subject
of recent discussions.

Another possible application  is    supersymmetry breaking in
low-energy supersymmetric models. Indeed, all models in which
supersymmetry breaking was obtained through a Fayet-Iliopoulos
mechanism were discarded ten years ago because the required
$U(1)$ symmetry was always anomalous. We have just shown that
many `a priori' anomalous $U(1)$ symmetries can be rescued
if anomalies are cancelled through a GS mechanism. It would be
interesting to check wether one can indeed build a realistic
supersymmetric Standard Model along these lines.

Finally, I have remarked           that in the absence of a GUT
symmetry (like e.g., in a hypothetical $SU(3)\times SU(2)\times U(1)$
string) there is no reason to insist in assuming the canonical
$k_1=5/3$ normalization. There could be other values (like
e.g. the example $k_1=3/2$ above) which may also lead to as good
        agreement with the low energy  measurements (or even better)
than the canonical one. Perhaps a reasonable procedure is the
opposite way around, assume unification at the Planck (or string)
scale and constraint the ratios $k_1/k_2, k_2/k_3$ from the
agreement with the low-energy data.

\bigskip
\bigskip
\leftline{\bf Acknowledgements}

I thank J.A. Casas, J. Louis, D. L\"ust, C. Mu\~noz,    F. Quevedo
and G.G. Ross for usefull discussions.

\endpage

\refout

\endpage

\bigskip
\bigskip

\begintable
   \
 \|         $Q$ | $u$     |$d$      |     $L$ |   $e$
\| ${\tilde H}$        |${\tilde {\bar H }}$
|   ${\tilde g},{\tilde W},{\tilde B}$                              \cr
     $R$    \| 0 | -1| 1 | 0 | 1 \| -1| 1 | 0              \cr
$A$  \|  0      |  0       |  -1     | -1      | 0       \| 1| 0  | 0 \cr
  $L_i$    \| 0        | 0        | 0   | -1 | 1 \|  0  |  0 |  0  \cr
  $P$        \|  -1  |  -1  |  -1  |  -1  |  -1  \|   1  |  1  |  1
\endtable

%\centerline{(a)}
\bigskip
\bigskip
\centerline{Table\ 1\ Generators of $U(1)$ symmetries in the SM and SSM}

\vfill\eject\bye